\documentclass[FBSedit,FBSmath,ecsub]{myFBSsuppl}
\usepackage{amsfonts}
\usepackage{amssymb}

\title{Does it make any sense to talk about a $\Delta$-isobar?}
\author{Frieder Kleefeld\thanks{\textit{E-mail address:} 
kleefeld@cfif.ist.utl.pt}}
\institute{Centro de F\'{\i}sica das Interac\c{c}\~{o}es Fundamentais, 
Instituto Superior T\'{e}cnico, Edif\'{\i}cio Ci\^{e}ncia, Piso 3, Av. Rovisco Pais, P-1049-001 LISBOA, Portugal}

\begin{document}

\maketitle
\begin{abstract}
It is shortly investigated, on what basis an experimentally observed resonance like the $\Delta (1232)$-isobar can be embedded into the framework of Quantum Theory (QT), i.e. Quantum Field Theory (QFT) and Quantum Mechanics (QM). After a short discussion of the particle concept in the context of the ``bootstrap'' idea of G.\ Chew and S.\ Mandelstam we will focus on the theoretical formalism being necessary to describe resonances in  the Lagrangian or Hamiltonian formulation of QT.
\end{abstract}

\section{Why this title?}
During a workshop on the (anticipated) occasion of the 65th birthday of Peter U. Sauer we want to recall the intriguing and revealing question raised by Peter U.\ Sauer \cite{gar1990} and collegues in the year 1990 on the resonance having excited him most of all: \textit{``Does it make any sense to talk about $N\Delta$ phase shifts?''}. In order to be allowed to ask this question we first have to explore the underlying question: \textit{``Does it make any sense to talk about a $\Delta$-isobar?''} or \textit{``Do unstable particles in physics have experimentally and theoretically a well defined meaning?''}.

The positive answer to the last question sketched in this short presentation will be based on previous work \cite{kle2002d,kle1999a,kle1998} (and references therein). Certain technical aspects will be displayed in more detail.

\section{Particle Concept and Bootstrap}
Experimentally it seems to be a not well defined problem to classify what is a physical ``particle'' and to specify most decisively its mass properties in physical data tables. The most accurate theoretical definition of a particle has been given by G.\ Chew (see ref. \cite{cap1}) who associated ``particles'' with poles of the scattering matrix. According to this definition the \textit{Review of Particle Properties} \cite{hagi2002} contains not only a small number of ``stable'', i.e.\ ``elementary'' particles (which play the quantum theoretical role of ``asymptotic states'') yet lists also a vast majority of ``particles'', which are not elementary. Such particles are either considered to be ``unstable'' or --- which is intimately related --- ``composite'' \cite{kle2002d,dass2002} (topical examples are e.g.\ scalar mesons \cite{kle2002c,kle2002b,bev2002}). As the parametrization of data in terms of (improved) non-relativistic Breit-Wigner fits being close to the idea of G. Chew fails in the description of processes of relativistic and strongly interfering unstable ``particles'' the developement of an adequate consistent relativistic formalism for such ``particles'' has been imperative. 
Consider a simple non-relativistic scattering problem at an energy at which the T-matrix can be described to a good approximation by a Breit-Wigner amplitude.
In this case the Breit-Wigner amplitude describes the full T-matrix already at ``tree-level''. The Breit-Wigner amplitude itself being determined by the solution of a Lippmann-Schwinger equation is perturbatively expanded into a Born series. Correspondingly, in the S-matrix picture of G. Chew the ``effective action'' which may be descibed by ``particles'' in the sense of G. Chew (which need not all to be stable) describes a scattering process already at ``tree-level'' while the ``effective action'' is usually perturbatively obtained from the generating functional of an interacting theory of asymptotic fields\footnote{Here we assume the existence of a Lagrangian and fields which is in general not guaranteed.}. The scenario that G.\ Chew's ``particles'' may be (at least approximately) chosen such that they describe scattering problems already (at least approximately) at ``tree-level'' may be summarized under the term ``bootstrap'' introduced by G.\ Chew and S.\ Mandelstam \cite{chew1961} (see also ref. \cite{cap1}). A nice feature of such quasi-bootstrap theories of (local) Chew ``particles'' is that even for large coupling constants physics is already exhaustively described at ``tree-level''\footnote{A typical example for a bootstrap theory is the ``quark-level linear sigma model'' (see e.g.\ ref.\ \cite{bev2002,sca2002} and references therein). Even for Hermitian fields nature is well described at tree-level and loop-corrections are estimated to be very small \cite{chan1973}.}. Yet there emerges the theoretically demanding consequence that not only self-energies, but also coupling constants may be complex valued. In order to handle such theories without getting in conflict with unitarity, causality, Lorentz covariance, locality, positivity, renormalizability etc.\ it is not only convenient, but imperative to achieve a consistent non-Hermitian prescription of QT in the context of a Lagrangian or Hamiltonian framework.
The solution to this problem has been illuminated to a certain extent in refs.\ \cite{kle2002d,kle1999a,kle1998}. It consists of the replacement of standard Hermitian QT by an (anti)causal formulation of QT, in which the underlying causal and anticausal Lagrangians (or Hamiltonians) are non-Hermitian. As a consequence the non-Hermitian causal and anticausal fields being solutions of the causal and anticausal non-Hermitian Lagrange equations of motion represent elements of a generalized (anti)causal, i.e. biorthogonal Fock space. This Fock space contains elementary (``quasi-Hermitian'') \textit{and} unstable (``non-Hermitian'') fields. As stated above we should bear in mind that the calculation of scattering or boundstate problems performed only in the ``stable'' particle basis is usually highly non-perturbative, while the inclusion of unstable particles might bring us by bootstrapping close to tree-level. 

\section{The (anti)causal Klein-Gordon (KG) field (``Nakanishi-field'')}
In order to tackle a spin-3/2-isospin-3/2-resonance like the $\Delta (1232)$ we first have to understand the properties of underlying (iso)spin 0, 1/2, 1 resonances. The (anti)causal spin 0 field has been studied for the first time\footnote{The considerations in the 2nd article of ref. \cite{kle2002d} have been done without knowing refs.\ \cite{nakanishi3,nakanishi2}.} by N.~Nakanishi  \cite{nakanishi3,nakanishi2} in the context of his ``Complex-Ghost Relativistic Field Theory'' defining a \textit{complex} (resonance) \textit{mass} $M=m-i\,(\Gamma/2)$ and denoting the ``free'' Lagrangian for the respective causal (anticausal) KG field $\phi_r(x)$ ($\phi_r^+(x)$) with isospin $(N-1)/2$ as ($r=1,\ldots,N$):
\[ {\cal L} (x) = \sum\limits_{r} 
\frac{1}{2} \Big\{ (\partial \phi_r (x) )^2  - M^2 (\phi_r (x) )^2 + 
 (\partial \phi_r^+ (x) )^2  - \, M^{\ast \, 2} (\phi_r^+ (x) )^2 \Big\}   \]
By variation of the action one obtains the respective causal and anticausal KG-equations $ (\,\partial^2 + M^2) \,\phi_r (x) = 0$ and $(\,\partial^2 + M^{\ast \, 2}) \,\phi_r^+ (x) = 0$, respectively.
As $\Im\,[M] = - \Gamma/2 \neq 0$ they are solved by a \textit{Laplace transform}\footnote{Let $\omega(\vec{p}):=\sqrt{\vec{p}^2 + M^2}$ ($\omega(\vec{0})=M$) and $a_r(\vec{p}):= a_r(p)|_{p^0 = \omega(\vec{p})}$, $c_r^+(\vec{p}):= a_r(- p)|_{p^0 = \omega(\vec{p})}$.}\comma{}\footnote{For details with respect to the Laplace transform involving a symbolic $\delta$-distribution denoted as ``$\delta (p^2 - M^2)$'' for the complex resonance mass $M$ see refs. \cite{nakanishi1,nakanishi2,kle2002d}!}: 
\begin{eqnarray} \phi_r (x) & = & \int \! \! \frac{d^4 p}{(2\pi )^3} \,\; \mbox{``} \, \delta(p^2 - M^2) \, \mbox{''} \,\; a_r \, (p) \, e^{ - i p x} \nonumber \\
   & \stackrel{!}{=} & \int \! \!
\frac{d^3 p}{(2\pi )^3 \; 2\, \omega (\vec{p})}
\;\;
\Big[ \, 
 a_r (\vec{p}) \; e^{ - i p x} +
 c_r^+ (\vec{p}) \; e^{+ i p x}
\,\Big] \Big|_{p^0 = \omega(\vec{p})}  \nonumber \\
\phi_r^+ (x) & \stackrel{!}{=} & \int \! \!
\frac{d^3 p}{(2\pi )^3 \; 2\,\omega^\ast (\vec{p})}
\;
\Big[\, 
 c_r (\vec{p}) \; e^{ - i p^\ast x} +
 a_r^+ (\vec{p}) \; e^{+ i p^\ast x}
\,\Big] \Big|_{p^0 = \omega(\vec{p})}  \nonumber
\end{eqnarray}
The Canonical conjugate momenta to $\phi_r (x)$ and $\phi_r^+ (x)$ are:
\[ \Pi_r (x) := \frac{\delta \,{\cal L} (x)}{\delta \, (\partial_0 \,\phi_r (x))}  \stackrel{!}{=} \partial_0 \,\phi_r (x) \; , \quad
 \Pi_r^+ (x) := \frac{\delta \,{\cal L} (x)}{\delta \, (\partial_0 \,\phi_r^+ (x))} \stackrel{!}{=} \partial_0 \,\phi_r^+ (x)  \]
The consistent (anti)causal quantization requires \textit{even} for \textit{resonance} fields \textit{equal-time} commutation relations. The respective non-vanishing relations are:
\[ { [ \, \phi_r (\vec{x},t) , \Pi_s (\vec{y},t) \, ] } = i\, \delta^{\, 3} (\vec{x} - \vec{y}\,) \; \delta_{rs} \; , \; 
{ [ \, \phi^+_r (\vec{x},t) , \Pi^+_s (\vec{y},t) \, ] }
 = i\, \delta^{\, 3} (\vec{x} - \vec{y}\,) \; \delta_{rs} \] 
Hence the non-vanishing commutation relations in momentum space are:
\begin{eqnarray} { [ \, a_r (\vec{p}) , c_s^+ (\vec{p}^{\,\prime}) \, ] } & = &
 (2\pi)^3 \, 2 \omega \,(\vec{p})\;\; \delta^{\, 3} (\vec{p} - \vec{p}^{\,\prime}\,) \; \delta_{rs}  \nonumber \\
 { [ \, c_r (\vec{p}) , a_s^+ (\vec{p}^{\,\prime}) \, ] } & = & (2\pi)^3 \, 2 \omega^\ast(\vec{p})\; \delta^{\, 3} (\vec{p} - \vec{p}^{\,\prime}) \; \delta_{rs} \nonumber 
\end{eqnarray}
The Hamilton operator is derived in the standard way from the Lagrangian\footnote{$H =  \int\! d^{\,3}x \;  \{ \sum_{r} 
( \Pi_r (x) \; (\partial_0 \,\phi_r (x))
+  (\partial_0 \,\phi^+_r (x)) \; \Pi^+_r (x) \,
) - \, {\cal L} (x) \}$}\comma{}\footnote{In 1-dim.\ QM we have $H = \frac{1}{2} \, \omega \, [c^+, a\, ]_\pm + \frac{1}{2} \, \omega^\ast \, [a^+, c\, ]_\pm$ ($\pm$ for Bosons/Fermions with $[c,a^+]_\mp =1$). This (anti)causal Harmonic Oscillator is diagonalized by the (normalized) right eigenstates $\left|n,m\right> = \frac{1}{\sqrt{n!\,m!}} \, (c^+)^n (a^+)^m \left|0\right>$ with $E_{nm} = \omega  \, (n \pm \frac{1}{2}) + \omega^\ast \, (m \pm \frac{1}{2})$.}:
\begin{eqnarray} H & = &  
\sum\limits_{r} \;\int\! d^{\,3}p \; \Big( \;
\frac{1}{2}\; \omega  (\vec{p})\; \Big\{c_r^+ (\vec{p}) , a_r(\vec{p}) \Big\} + 
\frac{1}{2}\; \omega^\ast (\vec{p})\; \Big\{ a_r^+ (\vec{p}) , c_r(\vec{p}) \Big\} \; \Big) \nonumber 
\end{eqnarray}
\textit{Standard real-time ordering} leads to the causal ``Nakanishi'' propagator:
\[ \left<0\right| T\,(\,\phi_r\,(x)\, \phi_s\, (y)\,) \left|0\right>  \stackrel{!}{=}  i \int\!\frac{d^{\,4}p}{(2\,\pi)^4}\;\; \frac{e^{-i\,p (x-y)}}{p^2 - M^2} \;\; \delta_{rs}  \]
The anticausal propagator is obtained by Hermitian conjugation or by a vacuum expectation value of an \textit{anti-real-time ordered} product of two anticausal fields.
\section{Lorentz boost, spinors and polarization vectors}
The definition of spinors and polarization vectors for spin 1/2 and 1 resonances requires the introduction of the respective concept of a Lorentz boost of unstable fields (see e.g.\ refs.\ \cite{kle2002d}). For any symmetric metric $g_{\mu\nu}$ Lorentz transformations  (LTs) $\Lambda^{\mu}_{\,\;\nu}$ are defined by $\Lambda^{\mu}_{\,\;\rho} \, g_{\mu\nu} \, \Lambda^{\nu}_{\,\;\sigma} = g_{\rho\sigma}$. Let $n^{\mu}$ be a timelike unit 4-vector $(n^2 = 1)$ and $\xi^{\mu}$ an \textit{arbitrary complex} 4-vector with $\xi^2 \not= 0$. The 4 independent LTs\footnote{They are well known as ortho-chronous/non-ortho-chronous proper/improper LTs.} relating $\xi^{\mu}$ with its ``restframe'' (i.e. $\xi^{\, \mu} = \Lambda^{\,\mu}_{\,\;\,\nu} (\xi) \; n^{\,\nu} \sqrt{\xi^2}$ and $n_{\,\nu} \sqrt{\xi^2} = \xi_{\, \mu} \, \Lambda^{\,\mu}_{\,\;\,\,\nu} (\xi)$) are ($P^{\,\mu}_{\,\;\,\nu} := 2\, n^{\,\mu} \, n_{\,\nu} - \, g^{\,\mu}_{\,\;\,\nu}$ = reflection matrix):
\begin{eqnarray} \Lambda^{\,\mu}_{\,\;\,\nu} (\xi) & = & \pm \, \left\{ g^{\,\mu}_{\,\;\,\rho} - \, \frac{\sqrt{\xi^2}}{\sqrt{\xi^2} \mp \xi \cdot n} \, \Big[ n^{\,\mu} \mp \frac{\xi^{\,\mu}}{\sqrt{\xi^2}} \Big] \Big[ n_{\,\rho} \mp \frac{\xi_{\,\rho}}{\sqrt{\xi^2}} \Big] \, 
\right\} \,  P^{\,\rho}_{\,\;\,\nu}  \nonumber \\
\Lambda^{\,\mu}_{\,\;\,\nu} (\xi) & = & \pm \, \left\{ g^{\,\mu}_{\,\;\,\nu} - \, \frac{\sqrt{\xi^2}}{\sqrt{\xi^2} \mp \xi \cdot n} \, \Big[ n^{\,\mu} \mp \frac{\xi^{\,\mu}}{\sqrt{\xi^2}} \Big] \Big[ n_{\,\nu} \mp \frac{\xi_{\,\nu}}{\sqrt{\xi^2}} \Big] \, 
\right\} \nonumber 
\end{eqnarray}
Let $\xi^\mu$ be e.g.\footnote{In ``Thermal Field Theory'' the inverse temperature $\beta = 1/T$ may be treated as an \textit{imaginary time}. Hence $\xi^\mu = (t - i\,\beta,\vec{x}\,)^\mu$ and $\sqrt{\xi^2} \, n^\mu = (\sqrt{(t - i\,\beta)^2-\vec{x}^{\;2}},\vec{0})^\mu$ are related by a LT!} the 4-momentum $p^\mu$ of a``force-free''\footnote{A remarkable ``bootstrap'' feature of (anti)causal QT is that a boost of such resonances does \textit{not} produce particles --- in contrast to the ``traditional'' boost of interacting particles!}  \textit{unstable} particle with complex mass $M$ in a $(+,-,-,-)$ metric. Its restframe is defined by $n^\mu = (1,\vec{0})^\mu$ \footnote{The space-time trajectories of such a particle are determined by the condition $p\cdot x = const$, or --- in 1 dimension --- by $x^1 = (\omega (\vec{p})/p^1) \, t + const$. If the movement proceeds along the \textit{real time axis}, the quantity  $\omega (\vec{p})/p^1$ has the interpretation of a \textit{complex} velocity.}. For such a spin 1/2 Fermion we define spinors $u(p) \equiv v(-p)$ and their transpose (i.e. $\overline{u^c}(p) := u^T(p) \; C$ with $C = i \,\gamma^{\,2} \, \gamma^{\,0}\,$) by the generalized Dirac equation $(\not\!p - \sqrt{p^2}\,)  u(p) = 0\,$ \footnote{Note that $u^c(p) = u(-\,p^\ast) = v(p^\ast)$. The spin projections are introduced such that for $\Re [p^0]\not=0$ there holds the spinor \textit{normalization} condition sgn($\Re [p^0] ) \, \sum_s u(p,s) \, \overline{v^c}(p, s) =  \; \not\!p +  \sqrt{p^2}$. The restframe 2-spinors used above obey $\chi_s^+ \chi_{s^\prime} = \delta_{s s^\prime}$ and $\sum_s \chi_s\chi^+_s= 1_2$.}. Lorentz covariance of this Dirac equation requires $S^{-1} (\Lambda (p)) \, \gamma^{\,\mu} \, S (\Lambda (p)) \, = \, \Lambda^{\,\mu}_{\,\,\;\nu} (p) \; \gamma^{\,\nu}$ for $u(p,s) = S(\Lambda(p)) \; u(\sqrt{p^2} \; n,s)$ resulting for the metric $(+,-,-,-)$ and the proper orthochronous LTs in
\begin{eqnarray} u(p,s) |_{\Re [p^0]>0} & = & \frac{\not\!p +  \sqrt{p^2}}{\sqrt{2\sqrt{p^2} \, (p^0 + \sqrt{p^2})}} \, u(\sqrt{p^2} \,n ,s) =  \frac{\not\!p +  \sqrt{p^2}}{\sqrt{p^0 + \sqrt{p^2}}} \left( \begin{array}{c} \chi_s \\ 0 \end{array}\right)  \nonumber \\
 \overline{v^c}(p,s) |_{\Re [p^0]>0} & = & \overline{v^c}(\sqrt{p^2} \,n ,s) \frac{\not\!p +  \sqrt{p^2}}{\sqrt{2\sqrt{p^2} \, (p^0 + \sqrt{p^2})}} = \left( \chi^+_s , 0 \right) \! \frac{\not\!p +  \sqrt{p^2}}{\sqrt{p^0 + \sqrt{p^2}}} . \nonumber
 \end{eqnarray}
The spin-projection vector $s^{\,\mu} (p) = \Lambda^\mu_{\;\nu} (p) \; s^{\,\nu} (\sqrt{p^2}, \vec{0}) = \Lambda^\mu_{\;\nu} (p) \, (0,\vec{s})^{\,\nu}$ with $|\vec{s}|^2=1$ obeys $s^{\,\mu} (p) s_\mu (p) = -1$ and $p^\mu  \, s_\mu (p) = 0$. For $\Re[p^0]>0$ there holds $\gamma_5 \not\!s (p) \, u(p,s) =  2 s \,\, u(p,s)$ and  $\gamma_5 \not\!s (p^\ast) \, v^c(p,s) =  2 s \,\, v^c(p,s)$ (with $s=\pm 1/2$). 
Similarly, defining 3 orthonormal Cartesian unitvectors $\vec{e}^{(i)}\; (i=x,y,z)$ by $\vec{e}^{(i)} \cdot \vec{e}^{(j)} = \delta^{ij}$ we may introduce polarization vectors by $\varepsilon^{\,\mu (i)} (p) := \Lambda^\mu_{\;\nu} (p) \; \varepsilon^{\,\nu \, (i)} (\sqrt{p^2},\vec{0}) = \Lambda^\mu_{\;\nu} (p) \, (0,\vec{e}^{(i)})^{\,\nu}$ fulfilling $\varepsilon^{\,\mu \, (i)} (p) \, \varepsilon_\mu^{\; (j)} (p) = - \,\delta^{ij}$, $p^\mu  \varepsilon_\mu^{\;(i)} (p) = 0$ and $\sum_i \varepsilon^{\,\mu \, (i)} (p) \, \varepsilon^{\,\nu \, (i)} (p) =  - g^{\,\mu\nu} + (p^{\mu} p^{\nu})/p^2$. All these identities can be projected on the complex mass shell $p^2 = M^2$, e.g. for $p^0 = \omega(\vec{p})$. 
\section{The (anti)causal Dirac  and Proca field}
As shown in refs.\ \cite{kle2002d,kle1999a,kle1998} the (anti)causal Lagrangian for an isospin $(N-1)/2$ spin 1/2 Fermionic resonance is given by ($r = 1,\ldots,N$)($\bar{M} := \gamma_0 \; M^+ \gamma_0$):
\[ {\cal L} (x) = \sum\limits_r \frac{1}{2} \, \Big( \overline{\psi_r^c} (x) \, ( \frac{1}{2} \, i \! \! \stackrel{\;\,\leftrightarrow}{\not\!\partial} \! -  M ) \, \psi_r (x) \; + \; \overline{\psi}_r (x) \, ( \frac{1}{2} \, i \!\! \stackrel{\;\,\leftrightarrow}{\not\!\partial} \! -  \bar{M} ) \, \psi_r^c (x) \; \Big) \]
As a whole we have four Lagrange equations of motion, i.e.  $( i \not\!\partial - M ) \, \psi_r (x) = 0$ and $( i \not\!\partial - \bar{M} ) \, \psi_r^c (x) = 0$ and the two \textit{transposed} equations for $\overline{\psi_r^c} (x)$ and $\overline{\psi_r} (x)$.
As for the KG field their solution is obtained by a \textit{Laplace transform} yielding\footnote{Here we have defined $q^\mu := (\omega(\vec{p}),\vec{p})^\mu = p^\mu |_{p^0 = \omega(\vec{p})}$ and  $b_r(\vec{p},s):= b_r(p,s)|_{p^0 = \omega(\vec{p})}$, $d_r^+(\vec{p},s):= b_r(- p,s)|_{p^0 = \omega(\vec{p})}$ and $u(\vec{p},s):= u(p,s)|_{p^0 = \omega(\vec{p})}$,  $v(\vec{p},s):= v(p,s)|_{p^0 = \omega(\vec{p})}$.}:
\begin{eqnarray} \psi_r (x) & = & \sum\limits_s \int \!\!\frac{d^3p}{(2\pi)^3 \, 2  \omega (\vec{p})} \,\; [ e^{- i  q x} b_{\,r} (\vec{p}, s) \, u(\vec{p},s)  + e^{i  q x} d^+_r (\vec{p}, s) \, v(\vec{p},s)] \nonumber \\
\psi_r^c (x) & = & \sum\limits_s \int \!\!\frac{d^3p}{(2\pi)^3 \, 2 \omega^\ast (\vec{p})} \, [ e^{i  q^\ast x} b^+_r (\vec{p}, s) \, u^c(\vec{p},s)  + e^{- i  q^\ast\cdot x} d_r (\vec{p}, s) \, v^c(\vec{p},s)] \nonumber \\
\overline{\psi_r} (x) & = & \sum\limits_s \int \!\!\frac{d^3p}{(2\pi)^3 \, 2 \omega^\ast (\vec{p})} \, [ e^{i  q^\ast x} b^+_r (\vec{p}, s) \, \bar{u}(\vec{p},s)  + e^{- i  q^\ast x} d_r (\vec{p}, s) \,\bar{v}(\vec{p},s) ] \nonumber \\
\overline{\psi^c_r} (x) & = & \sum\limits_s \int \!\!\frac{d^3p}{(2\pi)^3 \, 2 \omega (\vec{p})} \,\;\, [ e^{- i  q x} b_r (\vec{p}, s) \, \overline{u^c}(\vec{p},s)  + e^{i  q x} d^+_r (\vec{p}, s) \, \overline{v^c}(\vec{p},s)] \nonumber
\end{eqnarray}
Canonical (``\textit{real-equal time}'') quantization leads --- due to Fermi-statistics --- to the following non-vanishing momentum-space anticommutation relations:
\begin{eqnarray} { \{ \, b_r (\vec{p},s) , d_{r^\prime}^+ (\vec{p}^{\,\prime},s^\prime ) \, \} } & = &
 (2\pi)^3 \, 2 \,\omega \,(\vec{p})\;\; \delta^{\, 3} (\vec{p} - \vec{p}^{\,\prime}\,) \; \delta_{rr^\prime}\; \delta_{ss^\prime} \nonumber \\
{ \{ \, d_r (\vec{p},s) , b_{r^\prime}^+ (\vec{p}^{\,\prime},s^\prime ) \, \} } & = &
 (2\pi)^3 \, 2 \, \omega^\ast (\vec{p})\; \delta^{\, 3} (\vec{p} - \vec{p}^{\,\prime}\,) \; \delta_{rr^\prime}\; \delta_{ss^\prime} \nonumber
\end{eqnarray}
Convince yourself that --- as for the causal KG field --- the propagator of a causal spin 1/2 Fermion is obtained by \textit{standard Fermionic real-time ordering}\footnote{The anticausal propagator is obtained by Hermitian conjugation or by a vacuum expectation value of a \textit{anti-real-time ordered} product of two anticausal fields.}:
\[ \left<0\right| T\,(\,(\psi_r (x))_\alpha \, (\,\overline{\psi^c_s} (y))_\beta\,) \left|0\right>  \stackrel{!}{=} i \int\!\frac{d^{\,4}p}{(2\,\pi)^4}\;\; \frac{e^{-i\,p (x-y)}}{p^2 - M^2} \, (\not\!p + M)_{\alpha \beta}\; \delta_{rs} \]
As discussed in ref.\ \cite{kle2002d} we may use the formalism of Jun-Chen Su \cite{jun1} to introduce --- with the polarization vectors derived above --- a renormalizable and unitary Lagrangian for a massive (anti)causal vector field. In combining our results for a spin 0, 1/2 and 1 resonance the door of (anti)causal QT is opened widely for an (anti)causal $\Delta$-isobar.
\begin{acknowledge}
This work has been supported by the 
{\em Funda\c{c}\~{a}o para a Ci\^{e}ncia e a Tecnologia} \/(FCT) of the {\em Minist\'{e}rio da Ci\^{e}ncia e da Tecnologia (e do Ensinio Superior)} \/of Portugal, under Grant no.\ PRAXIS
XXI/BPD/20186/99 and SFRH/BDP/9480/2002.
\end{acknowledge}

\end{document}